\documentclass[10pt, twocolumn]{article}

\usepackage[margin=0.7in]{geometry}
\usepackage{graphicx}
\usepackage{xcolor}
\definecolor{receptorgray}{rgb}{0.95,0.95,0.95}
\usepackage{tikz}
\usepackage{mathtools}
\usepackage{amsmath}
\usepackage{makecell}
\usepackage[colorlinks,allcolors=blue]{hyperref}
\usepackage[font=footnotesize,labelfont=bf]{caption}

\usepackage[sorting=none,hyperref=true,style=phys]{biblatex}
\usepackage[switch]{lineno}

\newcommand{\receptor}[1]{
\tikz[baseline=0.2em]{\filldraw[color=#1,draw opacity=0] (0,0) rectangle (1em,1em)}
}
\newcommand{\kinase}[1]{
\tikz[baseline=-0.3em]{\filldraw[color=#1, draw opacity=0] (0,0) --(0.5em,0) circle (0.25em)}
}
\newcommand{\ecoli}{\textit{E.~coli }}

\begin{document}
\twocolumn[
\begin{center}
\textbf{\Large Lattice ultrasensitivity amplifies signals in \textit{E. coli}   without fine-tuning}\\
\vspace{10pt}
\small
Derek M. Sherry$^{1,2}$,
Isabella R.\ Graf$^{1,2,3}$,
Samuel J. Bryant$^{1}$,
Thierry Emonet$^{1,2,4}$,
Benjamin B. Machta$^{1,2}$
\\
\medskip
\noindent\upshape$^{1}$Department of Physics, Yale University, New Haven, Connecticut 06511, USA
\\
\noindent\upshape$^{2}$Quantitative Biology Institute, Yale University, New Haven, Connecticut 06511, USA
\\
\noindent\upshape$^{3}$Developmental Biology Unit, European Molecular Biology Laboratory, 69117 Heidelberg, Germany
\\
\noindent\upshape$^{4}$Department of Molecular, Cellular and Developmental Biology, Yale University, New Haven, Connecticut 06511, USA
\\
\noindent isabella.graf@embl.de
\\
\noindent thierry.emonet@yale.edu
\\
\noindent benjamin.machta@yale.edu
\vspace{10pt}
\end{center}]
\normalsize

\section*{Abstract}
The \textit{E. coli} chemosensory lattice, consisting of receptors, kinases, and adaptor proteins, is an important test case for biochemical signal processing.  Kinase output is characterized by precise adaptation to a wide range of background ligand levels and large gain in response to small relative changes in concentration. Existing models of this lattice achieve their gain through allosteric interactions between either receptors or core units of receptors and kinases. Here we introduce a model which operates through an entirely different mechanism in which receptors gate inherently far from equilibrium enzymatic reactions between neighboring kinases.
Our lattice model achieves gain through a mechanism more closely related to zero-order ultrasensitivity than to allostery. Thus, we call it lattice ultrasensitivity (LU).  Unlike other lattice critical models, the LU model can achieve arbitrarily high gain through time-scale separation, rather than through fine-tuning. The model also captures qualitative experimental results which are difficult to reconcile with existing models. We discuss possible implementations in the lattice's baseplate where long flexible linkers could potentially mediate interactions between neighboring core units.

\section*{Introduction}
The chemotaxis signaling pathway used by \textit{E. coli} for run and tumble navigation is perhaps the best understood biological circuit.  Predictive models can accurately describe how the sensory output -- activity of the kinase CheA -- responds to arbitrary time-varying inputs of ligand concentration \cite{sourjik.berg_2004,lan.etal_2011,tu_2013,goldman.etal_2009} either in bulk experiments \cite{sourjik.berg_2002,shimizu.etal_2010} or in individual cells \cite{keegstra.etal_2017,colin.etal_2017,kamino.etal_2020,moore.etal_2021,moore.etal_2024}. Striking features of the response are large gain and precise adaptation over a wide range of background concentrations; following a small step change in attractant concentration, CheA activity responds in a fast, highly amplified manner (large gain). Methyltransferase CheR and methylesterase CheB then methylate and demethylate receptors, slowly adapting the system output back to its original, steady-state level (precise adaptation). The proteins that implement these dynamics are well characterized and known structurally at atomic resolution~\cite{cassidy.etal_2023,djordjevic.stock_1997,djordjevic.etal_1998}. 
The minimal unit capable of responding to signals is called the core signaling unit (CSU), comprising one CheA homodimer, two trimers of transmembrane receptor dimers, and two copies of a structural protein called CheW~\cite{li.hazelbauer_2011,parkinson.etal_2015}. 
Neighboring CheAs and CheWs associate, connecting adjacent CSUs and forming the baseplate of a relatively stable lattice~\cite{briegel.etal_2009,liu.etal_2012,briegel.etal_2014}. 

Established models of chemosensing treat receptors as two state systems which completely determine the activity of their associated kinases \cite{sourjik.berg_2004,lan.etal_2011,tu_2013}. In these models, equilibrium allosteric coupling between receptors in the lattice allows for amplification of small signals while methylation allows for adaptation by shifting the relative free energies of the active and inactive state. While these models successfully capture the average response of signaling to time-varying ligand, they are hard to reconcile with some results.  First, experiments that specifically look at ligand binding find reduced gain and adaptation compared to kinase activity \cite{levit.stock_2002,amin.hazelbauer_2010}, inconsistent with the assumption of equilibrium interactions between receptors and kinases~\cite{hathcock.etal_2023}.  Second, recordings from individual cells show enormous fluctations~\cite{korobkova.etal_2004,keegstra.etal_2017,colin.etal_2017} and even coordinated two-state switching when receptors are homogeneous~\cite{keegstra.etal_2022}, strongly suggesting a model in which kinases are coupled through a lattice critical point~\cite{keegstra.etal_2022,hathcock.etal_2024}. Finally, experiments in mutants which fail to assemble individual core units into lattices also show reduced gain and adaptation~\cite{frank.etal_2016}.  While reduced gain is consistent with models in which allosteric interactions couple neighboring core units as in~\cite{hathcock.etal_2024}, such models do not easily account for reduced adaptation in these lattice deficient mutants.   

Motivated by these experimental results, we develop a model where gain arises primarily from chemical reactions carried out \textit{between} neighboring core units.  As with all previous models, receptors limit the activation and inactivation of individual kinases in a process which would --- in isolation --- yield a gain close to one. However, when core units are incorporated into lattices, receptors also gate non-equilibrium chemical reactions between kinases in neighboring core units in a process which naturally yields large gain. These reactions are far from equilibrium, facilitated by energy released during ATP hydrolysis. Rather than resulting from an allosteric mechanism, they likely involve chemical modification which could be mediated by the long flexible linker connecting the phosphoaccepting P1 domain to the rest of the kinase.  The gain in our model is mathematically equivalent to the zero-order ultrasensitivity described by Goldbeter and Koshland, wherein the concentration of the product of an enzymatic futile cycle depends sharply on the ratio of the rates of the forward and backwards enzyme reactions~\cite{goldbeter.koshland_1981}. For this reason, we call our model the Lattice Ultrasensitivity (LU) model.

We will use stochastic lattice simulations and an analytical mean-field approximation to explore the properties of the LU model.  We show that the LU model can capture experimental dose response curves similarly to classic models while also reproducing the large noise and two-state switching seen in measurements of single cell kinase activity~\cite{keegstra.etal_2017,colin.etal_2017}.  As with recent non-equilibrium models~\cite{hathcock.etal_2023,hathcock.etal_2024}, the LU model captures the differences in adaptation and gain between kinase activity and receptor binding assays. However, it also explains the reduced gain and adaptation in isolated core units.  
Perhaps most intriguingly, and different from all past lattice critical models, the LU model can achieve arbitrarily high gain through timescale separation without requiring any biologically encoded parameters to be fine-tuned. 

\section*{Model and Results}
\subsection*{Signaling in individual CSUs is enhanced through non-equilibrium coupling}
\begin{figure*}
    \centering
    \includegraphics[width=1\linewidth]{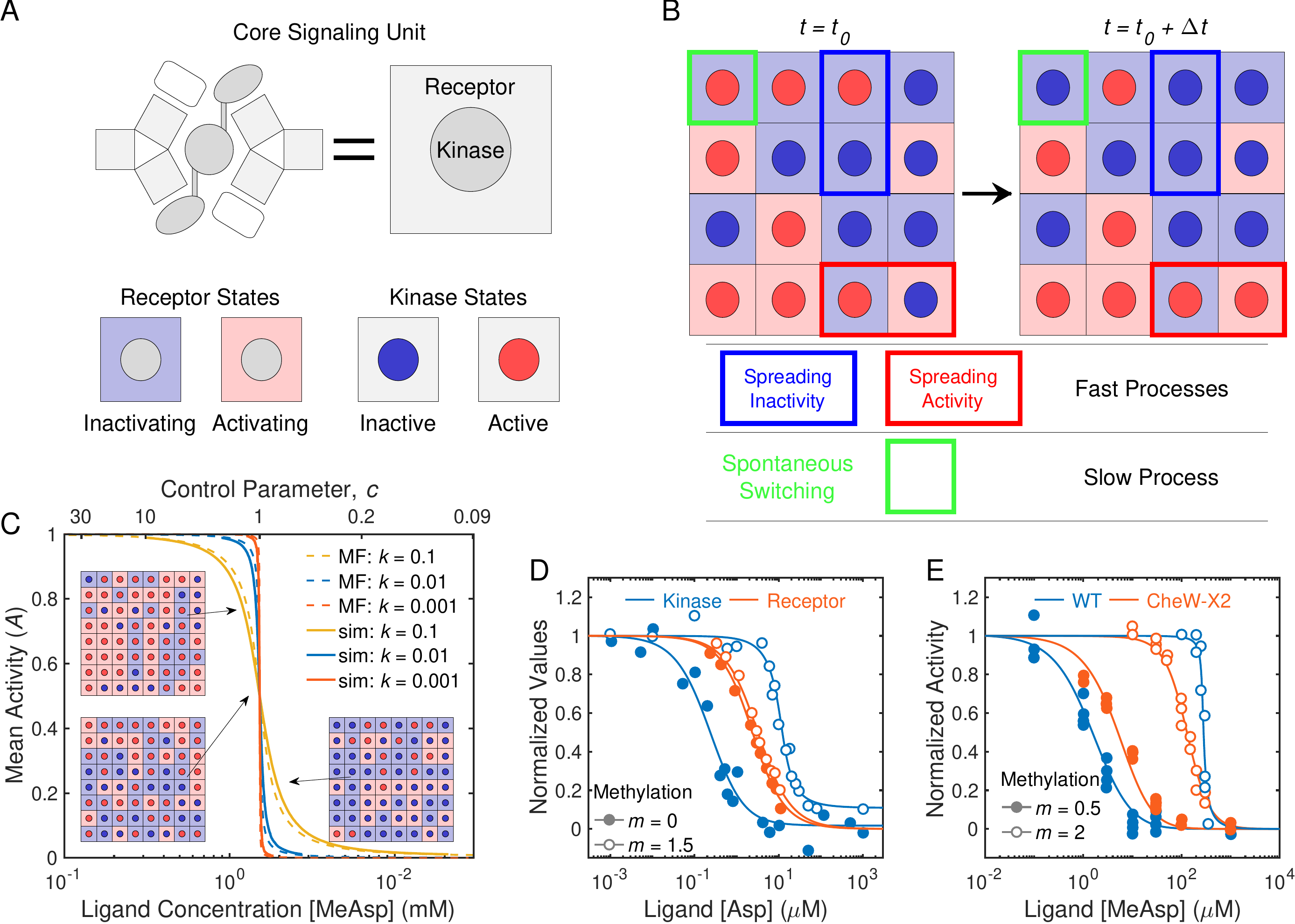}
    \caption{Overview of Lattice Ultrasensitivity.
    \textbf{A)} CheA kinases are represented by circles, whose color indicates their activity state (active or inactive). Receptors are represented by the surrounding squares, and their color indicates a preference for spreading of activity vs inactivity.
    \textbf{B)} Examples of spreading and spontaneous switching. The two configurations show one realization of each process occurring on a lattice in a time $\Delta t$. Colored rectangular outlines highlight the core units impacted by the processes indicated below. The probability of spreading activity or inactivity depends on the control parameter $c$, defined as the ratio of total rates of kinase activation and deactivation, which themselves depend on receptor occupancy and methylation (Eq. \ref{eq:c}). In this case $c=1$. 
    \textbf{C)} The cooperativity of dose response curves increases as $k$ becomes small. Mean field predictions agree with simulations. Insets show representative lattice configurations for the indicated activity levels; note that the changes in kinases activity between insets are amplified relative to changes in receptors. Here we've chosen $m = m_{0s}$.
    \textbf{D)} Fits to in-vitro experimental data from Amin and Hazelbauer~\cite{amin.hazelbauer_2010}. Nonequilibrium spreading allows for adaptation to be largely absent from receptor binding, but present in kinase activity.
    \textbf{E)} Separate fits to in-vitro experimental data from Frank et al.~\cite{frank.etal_2016}. When the sensory lattice is disrupted by the CheW-X2 mutant, spreading is no longer possible. Adaptation and gain are reduced, similarly to receptor binding.
    Fit parameters for (D) and (E) are in \ref{sec:fitting}.
    }
    \label{fig:overview}
\end{figure*}

We begin with a description of an individual, isolated CSU, containing a single receptor cluster and one kinase homodimer (Fig.~\ref{fig:overview}A). This CSU is described by three dynamic variables: the receptor state, the kinase activity, and the methylation level.

We model receptors in a CSU as an equilibrium Monod-Wyman-Changeux (MWC) cluster with a free energy difference between kinase-inactivating and kinase-activating states:
\begin{equation}
    \Delta f([L],m) = -\alpha_r(m-m_{0r}) +\mathrm{log} \left(\frac{1+[L]/K_i}{1+[L]/K_a}\right),
    \label{eq:deltaf}
\end{equation}
where $K_{i/a}$ are the ligand affinities of the two states, $\alpha_r$ parametrizes the dependence of the free energy on the methylation level $m$, $[L]$ is the instantaneous ligand concentration, and $\alpha_r m_{0r}$ is the free energy difference in the absence of ligand and methylation ($m=0$ and $[L]=0$).
The probability $p$ that a receptor is activating is then given by a Boltzmann distribution over the two states 
\begin{equation}
    p([L],m) =(1+e^{n\Delta f ([L],m)})^{-1},
\end{equation}
where $n$ denotes the degree of cooperativity within the receptor cluster. In standard MWC models for chemosensory gain, adaptation and gain are fully parameterized by $\alpha$ and $n$, respectively. In the LU model, these describe only the far smaller gain and methylation dependence observed in ligand binding assays~\cite{levit.stock_2002,amin.hazelbauer_2010}. Further gain and adaptation occur due to lattice interactions described below.

The kinase in a CSU can be active or inactive, and spontaneously switches between these two states at a base rate $k\nu/2$. Rather than directly controlling the kinase activity, the receptors gate transitions between activity states: a kinase can only become active if its associated receptor cluster is activating, and it can only become inactive if its associated receptor cluster is inactivating. 
No other dynamics are present in isolated CSUs. Thus, the average kinase activity exactly reflects the average state of the MWC-like receptors, and the activity of isolated CSUs exhibits the same reduction of gain and adaptation observed in receptor binding.

Enhanced gain and adaptation only appear in the LU model after CSUs are connected in an extended lattice where ``spreading'' reactions couple neighboring kinases. For simplicity we consider a square lattice (Fig.~\ref{fig:overview}B). An active kinase can spread its activity to a neighboring inactive kinase and activate it, but only if its neighbor's receptor is activating. This activation occurs at a methylation-dependent rate $\frac{\nu}{4} (1+e^{g(m)})$ per active kinase, where $\nu$ sets the base spreading rate independent of receptor state $p$ and methylation level $m$. Similarly, an inactive kinase can inactivate its neighboring active kinases at a rate $\frac{\nu}{4} (1+e^{-g(m)})$, but only if its neighbor's receptor is inactivating. 
Thus, the receptor state $p$ gates activity changes via spreading the same way they gate spontaneous switching.
We take $g(m) = \alpha_s (m - m_{0s})$, where $\alpha>0$ quantifies how strongly methylation affects spreading and $m_{0s}$ is the methylation level at which activation and inactivation rates are equal. 

We chose the specific form of methylation dependence so that the ratio of the rate of activation to deactivation is proportional to $e^{g(m)}$. This choice is consistent with other models \cite{shimizu.etal_2010,tu_2013,lan.etal_2011} where the linear $g(m)$ is the free-energy difference between active and inactive states. It is also consistent with experimental dose response curves (\ref{sec:fitting}) and ensures that the initial response speed after adaptation has minimal depdendence on ligand and methylation levels (\ref{sec:response}).

Taken together, all kinase and receptor dynamics in the LU model are summarized as follows:
\begin{align*}
    \receptor{{rgb:blue,2;white,3}} &\textrm{ : inactivating receptor} & \kinase{blue} &\textrm{ : inactive kinase}\\
    \receptor{pink} &\textrm{ : activating receptor} & \kinase{red} &\textrm{ : active kinase}\\
    \receptor{receptorgray} &\textrm{ : arbitrary receptor} & 
\end{align*}
\begin{align*}
    \receptor{pink}&\xrightleftharpoons[\gamma e^{-\Delta f([L],m)/2}]{\gamma e^{\Delta f([L],m)/2}}\receptor{{rgb:blue,2;white,3}} & &\genfrac{}{}{0pt}{}{\textrm{equilibrium receptor response}}{\textrm{(instantaneous)}(\gamma \gg \nu)}\\
    \mathrlap{\receptor{receptorgray}} \kinase{red} ~\mathrlap{\receptor{pink}} \kinase{blue} &\xrightarrow{\nu(1+e^{g(m)})/4} \mathrlap{\receptor{receptorgray}} \kinase{red} ~\mathrlap{\receptor{pink}} \kinase{red}  & &\textrm{activity spreading (fast)}\\
    \mathrlap{\receptor{receptorgray}} \kinase{blue} ~\mathrlap{\receptor{{rgb:blue,2;white,3}}} \kinase{red} &\xrightarrow{\nu(1+e^{-g(m)})/4} \mathrlap{\receptor{receptorgray}} \kinase{blue} ~\mathrlap{\receptor{{rgb:blue,2;white,3}}} \kinase{blue} & &\textrm{inactivity spreading (fast)}\\
    \mathrlap{\receptor{pink}} \kinase{blue} ~ &\xrightarrow{k\nu/2} \mathrlap{\receptor{pink}} \kinase{red} &  &\textrm{spontaneous activation (slow)}\\
    \mathrlap{\receptor{{rgb:blue,2;white,3}}} \kinase{red} ~ &\xrightarrow{k\nu/2} \mathrlap{\receptor{{rgb:blue,2;white,3}}} \kinase{blue} & &\textrm{spontaneous inactivation (slow)}
\end{align*}
$k$ is the ratio of switching to spreading when $g(m) = 0$. We will show that the LU model achieves large gain whenever $k\ll 1$, where spreading dominates kinase dynamics (Fig.~\ref{fig:overview}C). In this regime, low gain MWC-like receptors act as gates controlling spreading dynamics. By regulating the spread of activity rather than by coupling to it energetically, receptors need not demonstrate the gain and adaptation observed in assays of kinase activity (Fig.~\ref{fig:overview}D). And unique to our model, isolated core units not influenced by spreading dynamics show similarly reduced gain and adaptation (Fig.~\ref{fig:overview}E). We simulated these dynamics using a Gillespie algorithm \cite{gillespie_1977} on an extended lattice (\ref{sec:sim}), and the fitting procedure (\ref{sec:fitting}) used an analytical approximation described in the following section.

\subsection*{Mean-field dynamics explain the origins of large gain}
In order to understand how the dynamics achieve large gain, we develop an analytical mean-field theory, explicitly ignoring any spatial correlations that exist in the full system. In this approximation, the time evolution of the average kinase activity $A$ is described by a simple differential equation:
\begin{multline}
    \frac{1}{\nu}\frac{\mathrm{d}A}{\mathrm{d} t} = p([L],m) (1-A) \left((1+e^{g(m)})A+2k\right) \\
    - \left(1-p([L],m)\right)  A \left((1+e^{-g(m)})(1-A)+2k\right).
    \label{eq:mfDynamics}
\end{multline}
The first term on the right hand side is the rate at which inactive kinases becomes active via spreading and spontaneous switching, accounting for the restrictions placed on these processes by the receptor state. The second term similarly inactivation.

We start by examining the dynamics in the absence of switching ($k=0$), where the entire right hand side of Eq.~\ref{eq:mfDynamics} becomes proportional to $A(1-A)$. The right hand side is positive if $p([L],m) (1+e^{g(m)})>(1-p([L])) (1+e^{-g(m)})$ and negative otherwise. We will call the ratio of these two quantities the control parameter, 
\begin{equation}
\begin{split}
    c(m,[L]) &= \frac{p([L],m)(1+e^{g(m)})}{(1-p([L],m))(1+e^{-g(m)})} \\
    &= e^{g(m)-n\Delta f([L],m)},
\end{split}
\label{eq:c}
\end{equation}
which is an increasing function of $m$ and a decreasing function of $[L]$. In this $k\rightarrow0$ limit, when $c>1$ activity increases until it reaches a stable fixed point at $A=1$. When $c<1$ activity decreases and the stable fixed point is at $A = 0$. Thus, the steady state activity level is $A^* = \Theta(c-1)$, where $\Theta$ is the Heaviside function. At $c=1$, the susceptibility $dA/dc$ diverges and $A^*$ is extremely sensitive to small changes in $c(m,[L])$. 

Allowing for $k>0$, the steady state activity becomes a sigmoidal function of $[L]$ and $m$ (Eq.~\ref{eq:SA*}) which we can compare to experimentally measured dose response curves. In addition to predicting the receptor binding curves and CheW mutant activities shown earlier (Fig.~\ref{fig:overview}D,E), Eq.~\ref{eq:SA*} is consistent with dose response curves from a complete range of methylation levels \cite{shimizu.etal_2010}. Unless otherwise noted, we use the parameters found from fitting this last data set throughout the paper (Table \ref{tab:paramVals}). Details of the fitting procedure and parameters values for receptor binding and CheW mutant data sets can be found in \ref{sec:fitting}.

Intuition about the behavior of the sigmoidal dose response curve $A^*([L],m)$ can be gained by 
making a small modification to Equation~\ref{eq:mfDynamics}. If we multiply $k$ by $(1+e^{\pm g(m)})/2$, the methylation dependence of spreading and switching are identical, and the dependency of the steady state activity on $m$ and $[L]$ combine neatly into the control parameter $c(m,[L])$. We then find

\begin{equation}
    A^* = \frac{k(c+1)+1-c-\sqrt{(k(c+1)+1-c)^2-4ck(1-c)}}{2(1-c)}.
    \label{eq:A*}
\end{equation} 
When $m=m_{0s}$, $A^*$ here is exactly equal to the steady state $A^*([L],m)$ predicted by the unmodified dynamics (Eq.~\ref{eq:SA*}), and when $m\neq m_{0s}$ they remain qualitatively similar and have nearly identical gain (\ref{sec:aproxSS}). This suggests any intuition gained from the modified dynamics should still apply to the unmodified model.

Plotting Eq.~\ref{eq:A*} as a function of $c$ shows a sigmoid-like curve with inflection point at $c=1$ where there is precise symmetry between activity and inactivity (Fig.~\ref{fig:overview}C). As we reduce $k$, the curve becomes increasingly steep and switch-like, approaching the Heaviside function we found from our analysis as $k\rightarrow0$. Carefully taking the $k\rightarrow 0$ limit of Eq.~\ref{eq:A*}, we find that $\mathrm{d}A^*/\mathrm{d}c|_{c=1} = (8 k)^{-1}$. That is, for small $k$, the slope at the inflection point diverges like $1/k$. This divergence is the source of large gain in our model.

\begin{table}
\scriptsize
    \centering
    \begin{tabular}{c c c}
        \hline
        Parameter       & Description                                  & Value \\
        \hline
        $\nu$           & Base Spreading Rate                          & 100 s$^{-1}$ \\
        $k_R$           & Methylation Rate                             & 0.024 s$^{-1}$ \\
        $k_B$           & Demethylation Rate                           & 0.01 s$^{-1}$ \\
        $k$             & Ratio of Switching to Spreading              & 0.1 \\
        $n$             & Receptor MWC Factor                          & 2 \\
        $m_{0r}$        & Receptor Crossover Methylation Level         & -0.28 \\
        $\alpha_r$      & Receptor Free Energy per Methylation         & 0.99 kT \\
        $m_{0s}$        & Spreading Crossover Methylation              & 2.21  \\
        $\alpha_s$      & Spreading Effect per Methylation             & 0.92  \\
        $K_a$           & Dissociation Const. of Activating Receptor   & 474 $\mu$M \\
        $K_i$           & Dissociation Const. of Inactivating Receptor & 10 $\mu$M \\
        \hline
    \end{tabular}
    \caption{\textmd{Simulation Parameters. These values were used in all simulations unless otherwise specified. They were found by fitting data from~\cite{shimizu.etal_2010}. See \ref{sec:fitting} for details.}}
    \label{tab:paramVals}
\end{table}

The equation for the steady-state kinase activity $A^*$ (Eq.~\ref{eq:A*}), and that for the full model (Eq.~\ref{eq:SA*}), is identical to the fraction of modified substrate in Goldbeter and Koshland's model of zero-order ultrasensitivity, with the limit $k\rightarrow 0$ in our model corresponding to the case of precise enzyme saturation \cite{goldbeter.koshland_1981}.
While mathematically equivalent to zero-order ultrasensitivity, LU has a different structural origin. 
In each case, a forward reaction and a separate backward reaction work in opposition on the same substrate, and the system becomes ultrasensitive when the ratio of the forward to backward rate becomes relatively independent of the state of the substrate.  For zero-order ultrasensitivity, this relative independence occurs when the competing reactions work at saturation.
For the LU model, spreading of activity and inactivity both require a single active and inactive kinase to be neighbors on the lattice, so the ratio of their rates is independent of the average activity. The ratio of spontaneous activation and deactivation rates does depend on the average activity, but these processes are much slower than spreading, and are exactly analogous to a slight deviation from precisely balanced saturated enzymatic reactions in zero-order ultrasensitivity.

\subsection*{Fine-tuning is avoided through timescale separation and precise adaptation}
The LU model exhibits large gain whenever $k \ll 1$, but only if $c \approx 1$.
To robustly amplify signals in variable conditions, the system must maintain both parameters near their optimal values. 
The condition that $k \ll 1$ is not dynamically regulated in the LU model, but could be biologically implemented in a relatively straightforward manner through a separation of timescales between the rates of spreading and switching processes. Given such a separation of timescales, the precise adaptation of kinase activity known to exist in \textit{E. coli}~\cite{barkai.leibler_1997,alon.etal_1999} is sufficient to maintains the control parameter near its switch-like transition at $c=1$.

Precise adaptation in \textit{E. coli} is implemented by preferential methylation of receptors associated with inactive kinases and demethylation of receptors associated with active kinases ~\cite{morton-firth.etal_1999,parkinson.etal_2015}.
The LU model captures this phenomenology by assuming that the methylation rate of receptors takes the activity dependent form
\begin{equation}
    \frac{\mathrm{d} m}{\mathrm{d} t} = k_R (1-A) - k_B A^2,
    \label{eq:dmdt}
\end{equation}
where $k_R$ and $k_B$ are the rates of methylation and demethylation by CheR and CheB, respectively. The quadratic term for demethylation reflects the known phosphorylation of CheB by the kinase CheA.

\begin{figure*}
    \centering
    \includegraphics[width=1\linewidth]{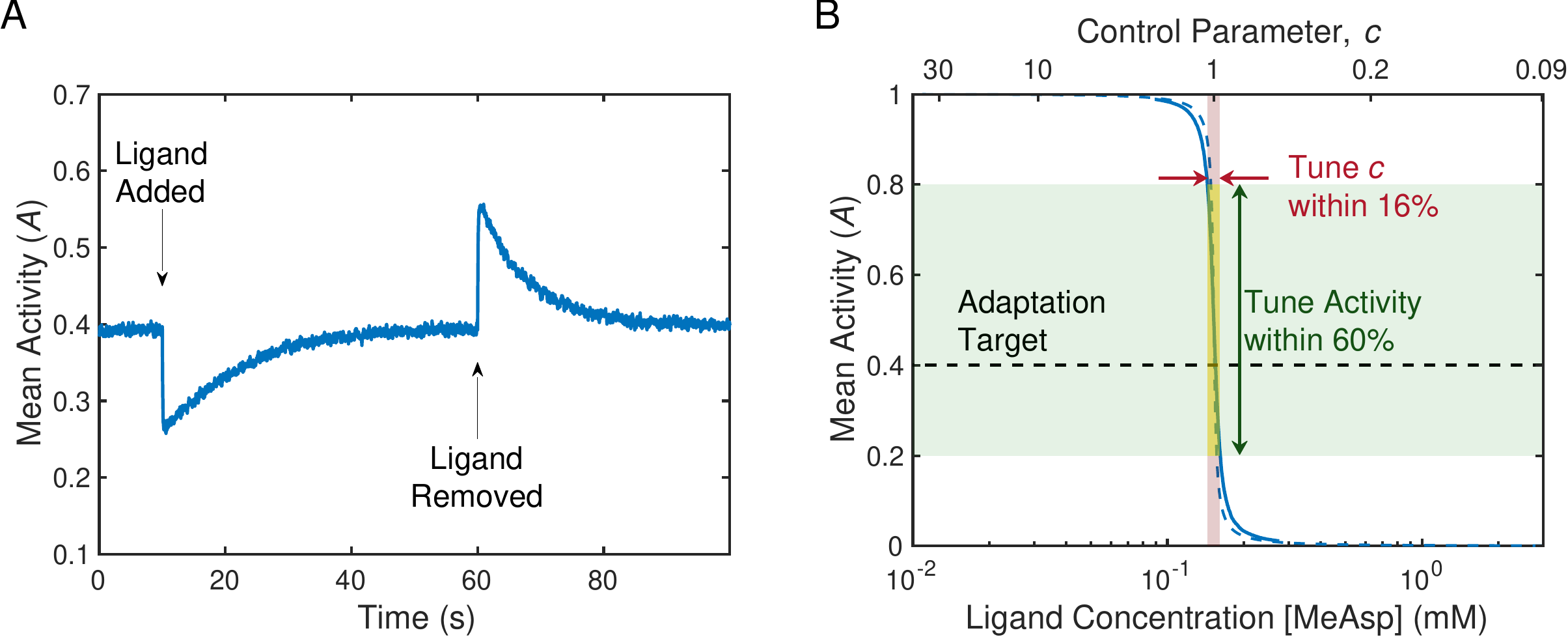}
    \caption{Adaptation in lattice ultrasensitivity.
    \textbf{A)} Methylation dynamics allow activity to precisely adapt after an initial fast response to stimulus. Simulations were initially adapted to 0.1 mM concentration of ligand. Plot shows the response to the addition and subsequent removal of 0.01 mM of ligand, averaged over 300 simulations.
    \textbf{B)} Because methylation is activity dependent and activity is a steep function of the control parameter, adaptation robustly maintains the system near $c=1$, ensuring large gain. If adaptation instead targeted the control parameter, it would be much less robust to variation. The 60\% range of the adaptation target is only an example to compare the relative robustness of tuning activity compared to the control parameter, not a specific value built into the model.
    For parameter values see Table~\ref{tab:paramVals}. All mappings from [MeAsp] to $c$ assume $m = m_{0s}$.} 
    \label{fig:adapt}
\end{figure*}

This is sufficient to ensure precise adaptation to some activity $0<\overline{A}<1$ (Fig.~\ref{fig:adapt}A). If both methylation and kinase activity are in steady state, then $\overline{A}=A^*$, which also corresponds to a steady state value of the control parameter $\overline{c}$. When $k\ll1$, $A^*$ approximates a step function of $c$, so a wide range of activity levels corresponds to a narrow range of $c$ near $c=1$. Thus, precise tuning of $\overline{A}$ via the adaptation parameters ($k_R$ and $k_B$) is not necessary to ensure that $\overline{c}\approx1$. As long as $k\ll1$, the methylation dynamics that achieve precise adaptation also ensure large gain without explicitly fine-tuning adaptation or control parameters (Fig.~\ref{fig:adapt}B).

These results for the LU model are reminiscent of feedback proposed in the context of self-organized criticality where the value of the order parameter, the collective output, feeds back onto the control parameter, thereby keeping the system close to its critical point~\cite{sornette_1992,sornette.etal_1995}.
In our model, this feedback is implemented by adaptation and keeps the control parameter near its critical value ($c=1$), as long as $k$ is small. The critical point in our model occurs at $k=0$, which is achievable by a separation of time-scales rather than by precise fine-tuning. This is similar to self-organized critical models that can be understood as remapping a system so that the critical point occurs when the control parameter is 0~\cite{sornette.etal_1995}.

\subsection*{Proximity to a critical point amplifies fluctuations and causes two-state switching}

\begin{figure*}[ht]
        \centering
        \includegraphics[width=1\linewidth]{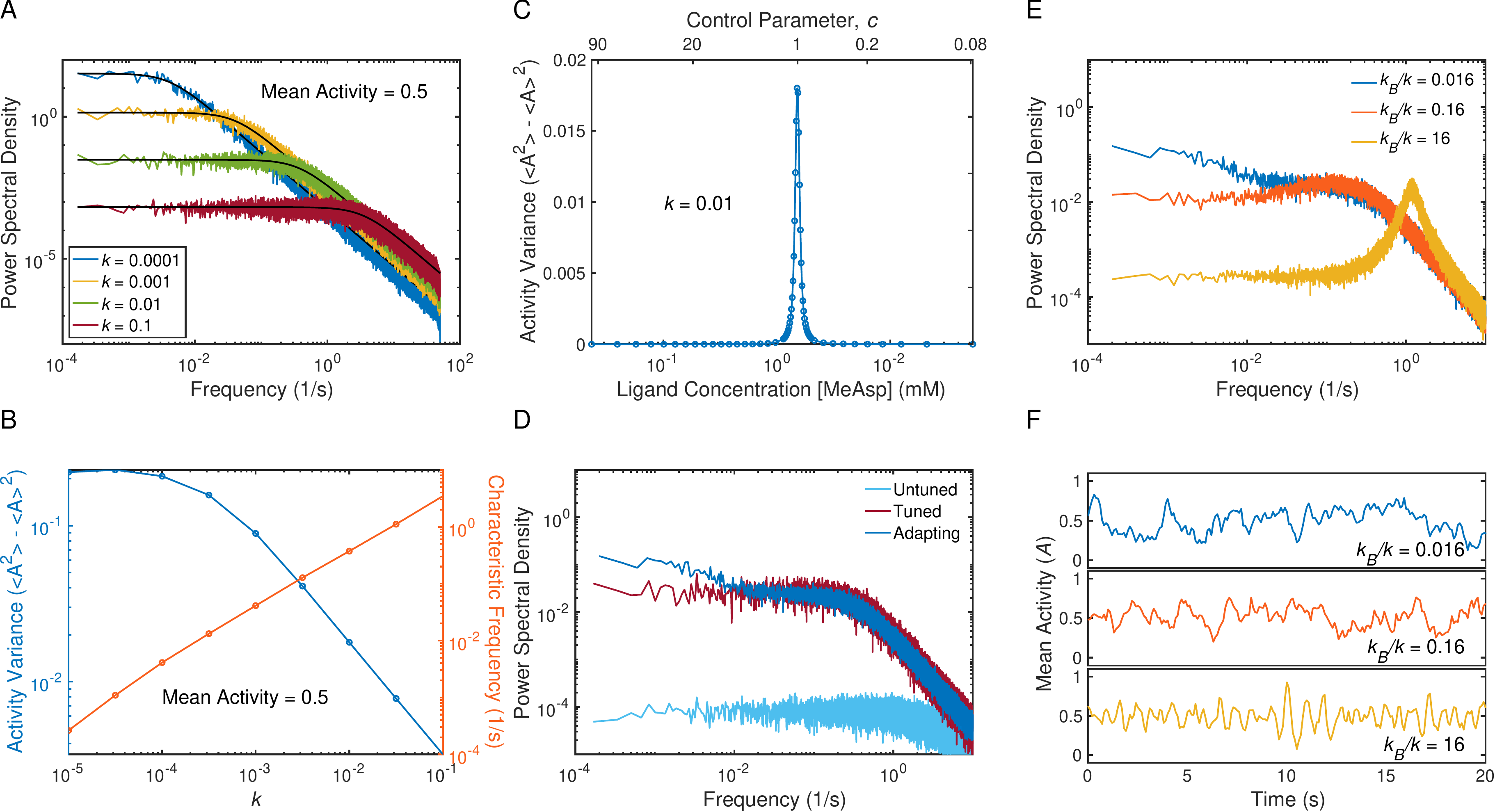}
        \caption{The LU model exhibits large fluctuations.
        \textbf{A)} Power spectral density of simulated kinase activity fluctuations at constant ligand concentration and methylation, for increasing values of $k$. Black lines show fits to a Lorentzian distribution.
        \textbf{B)} The characteristic frequency of fluctuations in nonadapting cells is linearly proportional to $k$. The variance shows power law scaling with $k$ when $k$ is large. When $k$ becomes sufficiently small, the variance saturates due to the onset of 2-state switching.
        \textbf{C)} For non-adapting cells, variance is sharply peaked near $c=1$ and small elsewhere.
        \textbf{D)} Power spectra show very small fluctuations in non-adapting cells without tuned ligand concentrations (mean activity $=0.1$). When cells have a control parameter close to 1 (mean activity $=0.5$), either through adaptation or tuning of ligand concentration, fluctuations are much larger.
        \textbf{E)} Three different regimes of adapting cells. $k_R/k_B$ is held constant throughout. When adaptation is much slower than $k$, two distinct timescales are visible in the power spectral density. When adaptation is faster than $k$, the PSD is peaked, indicating the presence of oscillation. Between these two limits, the PSD appears Lorentzian with a single timescale proportional to $k$.
        \textbf{F)} Example time series of cells corresponding to the three regimes depicted in E.
        For parameter values see Table~\ref{tab:paramVals}. All mappings from [MeAsp] to $c$ assume $m = m_{0s}$.}
        \label{fig:fluctuations}
    \end{figure*}

The LU model generates large activity fluctuations in proximity to its critical point. We simulated 100,000 seconds of activity in cells with different values of $k$ and fit the power spectra of their fluctuations to a Lorentz distribution (Fig.~\ref{fig:fluctuations}A). Methylation and ligand levels were held constant, so fluctuations were only due to stochastic kinase dynamics
as would be the case in $\Delta\textit{cheR}\Delta\textit{cheB}$ cells. As we reduced $k$ the characteristic timescale and variance of the fluctuations both increased, showing power-law scaling consistent with behavior near a critical point at $k=0$ (Fig.~\ref{fig:fluctuations}B). For small $k$, the size of the fluctuations is comparable to that of the entire system, consistent with single cells measurements~\cite{keegstra.etal_2017, colin.etal_2017}. The saturating variance seen for very small $k$ is because of this -- the finite lattice size means that once fluctuations encompass the entire system, they cannot grow any larger.

In agreement with experimentally observed fluctuations \cite{korobkova.etal_2004, keegstra.etal_2017, colin.etal_2017}, the LU model produces large fluctuations only when activity is brought to an intermediate level through adaptation or fine-tuning of ligand concentration. Indeed, the variance of the fluctuations is sharply peaked in the vicinity of $c=1$ (Fig.~\ref{fig:fluctuations}C). By bringing $c$ close to 1, adaptation thus leads to large fluctuations. To do this without adaption requires fine tuning the ligand concentration. This behavior is reflected in the power spectra. Adapting cells and non-adapting cells with finely tuned ligand concentrations have much larger fluctuations than non-adapting cells without fine tuning (Fig.~\ref{fig:fluctuations}D).

Adaptation dynamics also introduce noise via stochastic methylation, leading to small fluctuations in $c$ that change the shape of the power spectrum (Fig. {\ref{fig:fluctuations}E). When the sum of methylation rates, $k_B + k_R$, is much less than $k$, the system amplifies these fluctuations in $c$, adding a second, slower timescale to the power spectrum near the methylation frequency. When the sum of methylation rates is similar to $k$, the two timescales converge and the power spectrum again appears Lorentzian. If methylation rates increase further, adaptation is fast enough to respond to individual, stochastic changes in activity, suppressing low frequency fluctuations. However, it also allows for feedback between methylation and activity that leads to a high frequency peak in power spectrum and visible oscilations in the time series data (Fig.~\ref{fig:fluctuations}F). Experiments might be able to observe these oscillations by overexpressing adaptation enzymes in cells.

Two-state switching can occur in the LU model whenever fluctuations are large enough to push the system into a state where all the kinases are either active or inactive. Once the activity is uniform across the lattice, spreading can no longer occur. The lattice can only leave this uniform state via a spontaneous switching reaction. Because switching is much slower than spreading, once the lattice becomes uniformly active or inactive it will stay in that state for a relatively long time. This creates the appearance of two-state switching similar to what has been observed experimentally \cite{keegstra.etal_2017,keegstra.etal_2022}. 

Two simulation parameters control the relative size of fluctuations, and thus the presence of switching. Reducing $k$ increases the size of fluctuations, eventually leading to two-state switching (Fig.~\ref{fig:switching}A). Decreasing the lattice size can also drive switching (Fig.~\ref{fig:switching}B).

While internal dynamics are sufficient to produce fluctuations and two-state switching similar to experiments, sources of noise external to the LU model may also contribute to these behaviors. Indeed, the addition of noise to the transition rates in systems with zero-order ultrasensitivity has been shown to produce repeated transitions between bistable states~\cite{samoilov.etal_2005}. To investigate this possibility we introduced external noise by multiplying the spreading rate of activity by $e^{x_{OU}}$, where $x_{OU}$ is a variable obeying Ornstein-Uhlenbeck dynamics with zero mean, a defined variance, and a characteristic timescale. Increasing the variance of $x_{OU}$ made fluctuations larger, eventually producing two-state switching in simulations that would not otherwise have it (Fig.~\ref{fig:switching}C). Many known processes might contribute to external noise, such as the lattice exchanging receptors and kinases with its surrounding or performing dynamic structural rearrangements~\cite{frank.vaknin_2013}. Additionally, only about 15\% of measured cells exhibited two-state switching~\cite{keegstra.etal_2017}. If some of these sources of heterogeneity are themselves dynamic, they could also contribute to external noise.

\begin{figure*}
    \centering
    \includegraphics[width=1\linewidth]{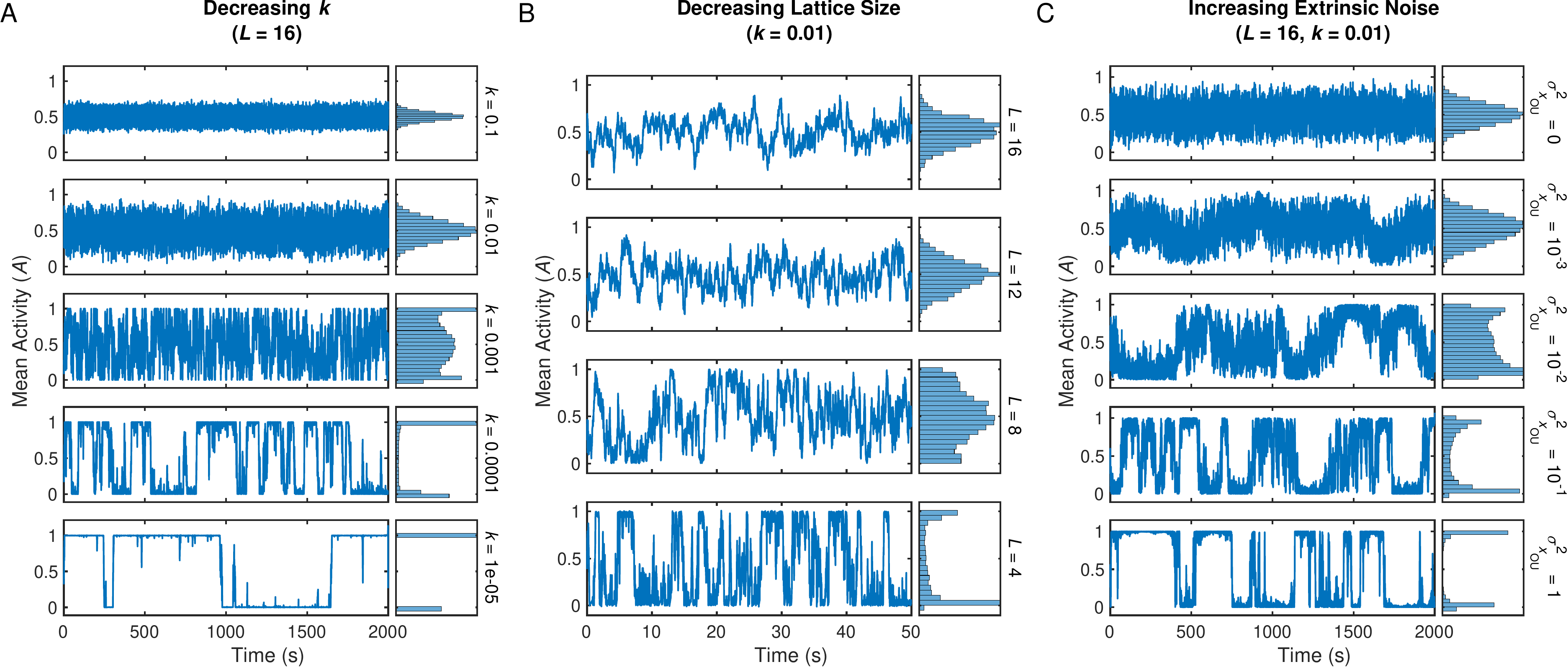}
    \caption{Emergence of two-state switching. 
    \textbf{A)} Simulated time series with mean activity set to 0.5 and decreasing $k$ from top to bottom. As $k$ decreases, fluctuation size and timescales grow. Once fluctuations are large enough to reach the boundaries of the system ($A$ = 0 or $A$ = 1), switching begins to occur. 
    \textbf{B)} As in (A), but now decreasing the linear dimension of the lattice, $L$, from top to bottom. As $L$ decreases, the size of fluctuations relative to the total number of kinases increases. Switching occurs when the system is small enough that the fluctuations reach the boundaries. The x-axis is zoomed in to make high frequency switching clear.
    \textbf{C)} As in (A), but we have introduced noise into the control parameter. Variance of the noise increases from top to bottom. 
    Sufficiently large noise in the control parameter induces two-state switching in what would otherwise be a non-switching system. Here $k$ = 0.01, equivalent to the second plot in (A). 
    For parameter values see Table~\ref{tab:paramVals}.}
    \label{fig:switching}
\end{figure*}

\section*{Summary and Discussion}
We have presented a novel lattice model for the large gain observed in \textit{E. coli} chemosensing. The LU model uses non-equilibrium reactions between CSUs to explain the large gain and adaptation seen in wild-type cell's kinase activity.  Our model can also account for the reduced gain and adaptation seen in lattice mutants and in receptor binding assays, and it recapitulates the large fluctuations and two-state-switching seen in single cell experiments. Finally, unlike other lattice critical models, the LU model does not require any fine-tuned parameters.

The critical point in our model reproduces the most salient features of fluctuations measured in single cells: large slow fluctuations, and two-state switching in a subset of non-adapting cells~\cite{keegstra.etal_2017, colin.etal_2017, keegstra.etal_2022}. The magnitude of these fluctuations is larger than expected in MWC-like models where kinases act independently or in small groups. Indeed, the fluctuations and two-state switching are difficult to reconcile with any model without a lattice critical point. 
Both our model and Ising models contain such a critical point. However, the LU model achieves that critical point without fine-tuning coupling parameters, while the Ising model requires its coupling constant to be close to a specific finite value.

Other biological sensory systems also exploit the divergent susceptibility found near a critical or bifurcation point to amplify weak signals. 
Hair cells amplify faint sounds in the inner ear by tuning it close to a Hopf bifurcation~\cite{camalet.etal_2000, choe.etal_1998, eguíluz.etal_2000, hudspeth.etal_2010, momi.etal_2025}. We have argued that pit vipers use a saddle-node bifurcation to amplify milli-Kelvin changes in the temperature of their sensory organ into a robust neural response ~\cite{graf.machta_2024}, and that olfactory receptor neurons in fruit flies exploit a bifurcation to extract odor timing information for navigation~\cite{choi.etal_2024}.
In all three cases, it has been argued that order-parameter based feedback, originally introduced in the context of self-organized criticality~\cite{sornette_1992, sornette.etal_1995} and bistability~\cite{disanto.etal_2016, buendía.etal_2020}, allows the system to maintain proximity to the critical or bifurcation point without the need for fine-tuning.
In our model, this feedback from the order parameter back onto the control parameter is implemented by the activity dependence of methylation dynamics~\cite{barkai.leibler_1997, yi.etal_2000, aoki.etal_2019}.  

That coupling between CSUs in the LU model is explicitly out of equilibrium and distinguishes it from other models. The enhanced gain and adaptation measured in kinases relative to receptors~\cite{levit.stock_2002, amin.hazelbauer_2010} is captured by the LU model (Fig.~\ref{fig:overview}D) but cannot be explained by equilibrium models \cite{hathcock.etal_2023}. Other models have also explained this by introducing non-equilibrium interactions within CSUs~\cite{hathcock.etal_2023, hathcock.etal_2024}. However, certain CheW mutants show a reduction in adaptation and gain similar to receptors~\cite{frank.etal_2016}, suggesting that by disrupting connections between CSUs these mutants also disrupt the non-equilibrium process responsible for enhancing gain and adaptation. This implies that the non-equilibrium process involves the coupling between neighboring CSUs as in the LU model, not just interactions within individual CSUs.

The mechanism responsible for non-equilibrium coupling (i.e spreading) between CSUs should be quite different from the allosteric interactions assumed to explain previous equilibrium couplings. One  possible mechanism involves P1, the phospho-accepting domain of CheA. 
P1 is connected to the rest of the CheA dimer by an extended flexible linker. This linker allows P1 to be phosphorylated by the P4 domain of the opposite CheA in the dimer pair, but it is long enough that it might facilitate reactions with neighboring core units as well~\cite{burt.etal_2021}. Thus, spreading could occur via direct phosphotransfer reactions between neighboring CheA mediated by P1.

The molecular implementation of spreading is poorly constrained due to the large number of states and interactions described among the proteins in the signaling lattice. However, future experiments may be able to improve the situation. For instance, P1-mediated interaction could be tested by shortening the flexible linkers, making interactions between neighboring kinases less likely. This could reduce gain, but might entirely stop signal transduction first. As for lengthening linkers, experiments have already shown that mutants whose P1 and P2 domain are entirely separate from the lattice and soluble in the cytoplasm are still able to chemotax~\cite{garzón.parkinson_1996}, but quantitative probes of signaling activity in these cells have not been performed. If the reduced gain and adaptation seen in CheW mutants \cite{frank.etal_2016,piñas.etal_2016} is because core units are too spatially separated for their linkers to reach each other, perhaps mutants with soluble P1-P2 domains can partially rescue large gain in the CheW mutants. Intriguingly \textit{Vibrio cholerae} has far fewer and therefore sparser CheA \cite{yang.etal_2018}, but their linker is significantly longer \cite{jumper.etal_2021,varadi.etal_2022,varadi.etal_2024}, potentially allowing CheAs to interact over longer distances. 

Other experiments may be able to probe different features of the LU model. The LU model predicts that balanced over-expression of adaptation enzymes CheR and CheB should lead to oscillations in kinase activity (Fig.~\ref{fig:fluctuations}E,F), although this prediction may not be unique to the LU model. Additionally, the non-equilibrium nature of the proposed spreading interaction means that experiments restricting the free energy available via ATP hydrolysis should meaningfully affect signaling. Further theoretical work is needed to clarify precisely how limiting available energy would impact measurable signaling characteristics such as response time, gain, and variance.

The LU model presents a qualitatively different mechanism for achieving sensory gain through a lattice critical point. LU dynamics are driven by directed chemical reactions between neighboring signaling enzymes rather than by thermal fluctuations.  The result is a mechanism for amplification that resembles zero-order ultrasensitivity.  This mechanism seems to be more robust- acting as an effective amplifier without the finely tuned energetic coupling that is required for analogous equilibrium critical points.  While coarse-grained, the LU model provides new ideas for interpreting the structure, function and design principles of receptors and kinases in the chemosensory array.

\section*{Acknowledgements}
We thank Jeremy Moore, Swayamshree Patra, Tom Shimizu, and the members of the Machta group for helpful discussions, and Michael Abbott, Johannes Keegstra, Asheesh Momi, and Mason Rouches for constructive comments on the manuscript. This work was supported by the Sloan Alfred P. Foundation award G-2023-19668 (TE,BBM), by the Deutsche Forschungsgemeinschaft (DFG, German Research Foundation) – Projektnummer 494077061 (IRG), by NIH R35GM138341 (BBM), and by NIGMS awards GM106189 and GM138533 (TE).}

\printbibliography
\onecolumn
\begin{center}
\textbf{\LARGE Supplemental Material to\\ Lattice ultrasensitivity produces large gain in \ecoli chemosensing} \\
\vspace{10pt}
\large
Derek M. Sherry$^{1,2}$,
Isabella R.\ Graf$^{1,2,3}$,
Samuel J. Bryant$^{1}$,
Thierry Emonet$^{1,2,4}$,
Benjamin B. Machta$^{1,2}$
\\
\medskip
\noindent\upshape$^{1}$Department of Physics, Yale University, New Haven, Connecticut 06511, USA
\\
\noindent\upshape$^{2}$Quantitative Biology Institute, Yale University, New Haven, Connecticut 06511, USA
\\
\noindent\upshape$^{3}$Developmental Biology Unit, European Molecular Biology Laboratory, 69117 Heidelberg, Germany
\\
\noindent\upshape$^{4}$Department of Molecular, Cellular and Developmental Biology, Yale University, New Haven, Connecticut 06511, USA
\\
\end{center}
\setcounter{equation}{0}
\setcounter{figure}{0}
\setcounter{table}{0}
\setcounter{section}{0}
\makeatletter
\renewcommand{\theequation}{S\arabic{equation}}
\renewcommand{\thefigure}{S\arabic{figure}}
\renewcommand{\thesection}{S\arabic{section}}
\renewcommand{\thetable}{S\arabic{table}}

\large

\section{Determining Parameter Values}
\label{sec:fitting}
\subsection{Fitting Dose Response Curves}
For each data set that we fit, we used the same general procedure. Each dose response curve in the data set was normalized by its mean value, and data corresponding to specific methylation mutants were assigned a specific value of $m$ for fitting (EEEE: $m=0$, QEEE: $m=0.5$, QEQE: $m=1$, QQQE: $m=1.5$, QQQQ: $m=2$, QEmQQ: $m=2.5$, QEmQEm: $m=3$, EmEmEmEm: $m=4$). All curves in a given data set were then simultaneously fit to the mean-field equation for the LU model using MATLAB's nonlinear least-squares curve-fitting function.

Except for CheW mutants, all kinase activity dose response curves were fit to the steady state activity for the full model
\begin{equation}
    A^* = \frac{1-c+\frac{2k}{1+e^{g(m)}} + \frac{2 c k}{1+e^{-g(m)}}-\sqrt{(1-c+\frac{2k}{1+e^{g(m)}}+\frac{2 c k}{1+e^{-g(m)}})^2 - 4(1-c)(\frac{2 c k}{1+e^{-g(m)}})}}{2(1-c)},
    \label{eq:SA*}
\end{equation}
with all terms defined in the main text. Assuming that the rate of spreading in CheW mutants is zero, resolving the dynamical equations reveals that the steady state activity of these mutants is just
\begin{equation}
    A^*_{CheW}=p([L],m).
\end{equation}
Based on the differing ligand affinities of the activating and inactivating receptors, the equations we fit to the ligand binding curves was
\begin{equation}
    b = \frac{L}{K_i+L}(1-p([L],m))+ \frac{L}{K_a+L}(p).
\end{equation}

\begin{table}
    \centering
    \begin{tabular}{c c c c} 
        \hline
        Parameter       & Description                                  & \makecell{Values for in-vitro\\ receptor binding~\cite{amin.hazelbauer_2010}}     & \makecell{Values for in-vivo\\ CheW mutants~\cite{frank.etal_2016}}\\
        \hline
        $k$             & Ratio of Switching to Spreading              & 0.1            & 0.1\\
        $n$             & Receptor MWC Factor                          & 2              & 2   \\
        $m_{0r}$        & Receptor Crossover Methylation Level         & -3.35          & 0.69 \\
        $\alpha_r$      & Receptor Free Energy per Methylation         & 0.34 kT        & 2.09 kT\\
        $m_{0s}$        & Spreading Crossover Methylation              & 1.72           & -0.33\\
        $\alpha_s$      & Spreading Effect per Methylation             & 1.57           & 0.56\\
        $K_a$           & Dissociation Const. of Activating Receptor   & 4.5 $\mu$M     & 126 mM\\
        $K_i$           & Dissociation Const. of Inactivating Receptor & 0.85 $\mu$M    & 9 $\mu$M\\
        \hline
    \end{tabular}
    \caption{Parameters found from fitting data from Ref.~\cite{amin.hazelbauer_2010} and Ref.~\cite{frank.etal_2016}.}
    \label{tab:SparamVals}
\end{table}



\begin{figure}
    \centering
    \includegraphics[width=0.6\linewidth]{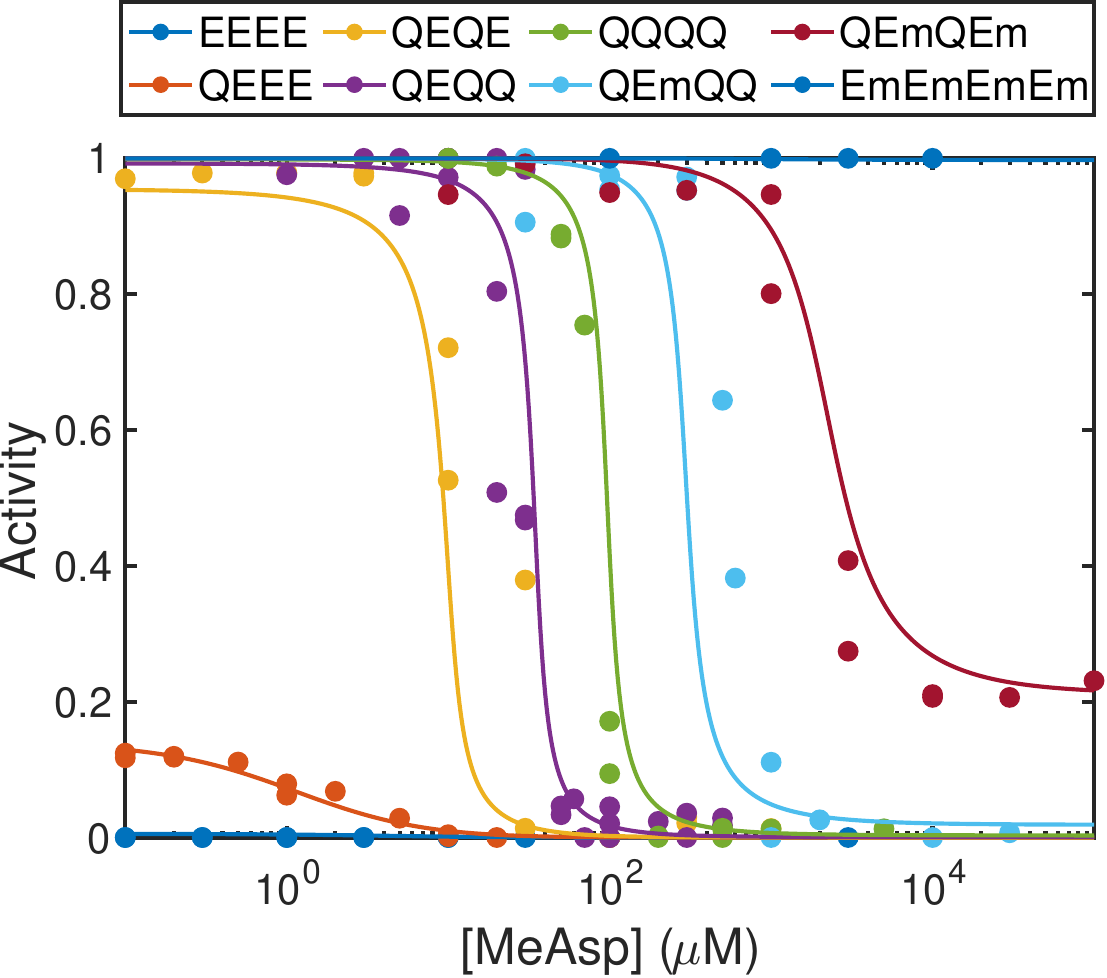}
    \caption{Fits to experimental dose response curves from Ref.~\cite{shimizu.etal_2010}.} 
    \label{fig:DRCfit}
\end{figure}

\subsection{Finding the remaining parameters}
After fitting the dose response curves, we still need to determine the base spreading rate $\nu$ and the methylation and demethylation rates $k_R$ and $k_B$.
The steady state activity level after adaptation, $\bar{A} = 2/(1+\sqrt{1+4k_B/k_R})$, determines the ratio $k_R/k_B$. We chose $\bar{A} = 0.3$ in rough agreement with the average of cells from Ref.~\cite{mattingly.etal_2021}, corresponding to $k_R/k_B=2.4$.
The total of the (de)methylation rates, $k_R+k_B$, determines the adaptation timescale, which is roughly 10 seconds \cite{mattingly.etal_2021}. After simulating the response of adapting cells to a 10\% change in ligand concentration for a variety of values, we found that $k_R+k_B = 0.034 \mathrm{s}^-1$ roughly corresponds to an adaptation timescale of 10 s. Because of significant variability between individual cells, a more precise fitting procedure for the (de)methylation rates would provide meaningfully better information.

Because the response timescale of kinase activity is smaller than the time resolution available from FRET microscopy \cite{mattingly.etal_2021}, it cannot be effective constrained by FRET measurements. This, in turn, means that the base spreading rate $\nu$ is also poorly constrained, with the only requirement being that our system responds faster than the resolution of FRET microscopy, roughly 500 ms. We chose $\nu = 100~\mathrm{Hz}$, comfortably above this minimum rate.

These additional parameters are summarized in Table~\ref{tab:SparamVals2}.

\begin{table}
    \centering
    \begin{tabular}{c c c} 
        \hline
        Parameter       & Description                                  & Value \\
        \hline
        $\nu$           & Base Spreading Rate                          & 100 s$^{-1}$ \\
        $k_R$           & Methylation Rate                             & 0.024 s$^{-1}$ \\
        $k_B$           & Demethylation Rate                           & 0.01 s$^{-1}$ \\
        \hline
    \end{tabular}
    \caption{Additional parameter values chosen for use in our model.}
    \label{tab:SparamVals2}
\end{table}

\section{Description of Simulations}
\label{sec:sim}
\subsection{Setup and Initialization}
In our simulation, core signaling complexes (CSCs) are located on the sites of a square lattice. There are two values associated with each CSC on the lattice: a kinases activity ($A$), and a methylation level ($m$). The kinase activity takes a value of 0 or 1, corresponding the inactive or active state, respectively. The methylation level takes a value between 0 and 4. Since there are 12 receptors (2 trimers of dimers) in each CSC, we discretize methylation levels into steps of $dm = 1/12$, corresponding to the methylation of a single site of one receptor. A methylation level of 0 or 4 means that every receptor in the CSC is either fully demethylated or methylated. There is also a value $p$ associated with the state of the receptor in each CSC. We assume that the dynamics of the receptor state are in fast equilibrium relatively to the other dynamics of the system, so rather than take a value of 0 or 1 corresponding to the inactivating or activating states, it takes the average state, $p$, somewhere between 0 and 1.
\begin{align}
    p([L],m) = (1+e^{n\Delta f ([L])})^{-1}.
    \label{eq:Sunbound_fraction}
\end{align}
\begin{equation}
    \Delta f([L],m) = \Delta f_0(m) +\mathrm{log} \left(\frac{1+[L]/K_i}{1+[L]/K_a}\right)
    \label{eq:Sdeltaf}
\end{equation}
Because this model only accounts for a single type of receptor, and all the receptors are in fast equilibrium, we treat the value of the receptor state as a deterministic global variable rather than a stochastic variable with a different value at each CSC.

To initialize the simulation with adaptation, the kinase activity and methylation level of each CSC is determined randomly. The receptor state is determined based on the initial ligand concentration. If we are simulating an non-adapting mutant (e.g. QEQE), initialization of the kinase activity and the receptor state occurs as above, but the methylation level of each CSC is set to a constant value that is determined by the mutant in question.

\subsection{Dynamics}
Dynamics are then simulated using a Gillespie algorithm \cite{gillespie_1977} with unique reactions for each CSC. The propensity for reactions at the $i$th CSC are given by the following: An inactive kinase becoming active 
\begin{equation}
   r_{0\rightarrow1}^{A_i} = p \left(k + \frac{1 + e^{g(m)}}{4}\sum_{j \in n.n.} A_j\right),
\end{equation}
where $m$ is the methylation level of the CSC, other variables are defined as in the main paper, and the sum is over the activity values of nearest neighbor kinases.
The propensity for active kinases to become inactive is given by 
\begin{equation}
    r_{1\rightarrow0}^{A_i} = (1-p)\left((k+(1 + e^{-g(m)})(1-\frac{1}{4}\sum_{j \in n.n.} A_j)\right).
\end{equation}
The propensity to increase the methylation by $dm$ is given by 
\begin{equation}
    r^{m_i}_{+dm} = k_R(1-A_i).
\end{equation} 
The propensity to decrease the methylation by $dm$ is given by 
\begin{equation}
    r^{m_i}_{-dm} = k_B A_i \frac{1}{N}\sum_{j=1}^N A_j.
\end{equation}
Here, the sum is over every lattice site, making the demethylation rate proportional to both the single site activity and the average over the entire lattice. This accounts for the activation of CheB via phosphorylation by CheA.

Sometimes, rather than simulate the lattice at a specific methylation level and/or ligand concentration, it is convenient to just set the control parameter $c = e^{m_0 + \alpha m_i} p/(1-p)$, which combines the influence of methylation and ligand into a single parameter. In this case, the propensity for methylation and demethylation are unchanged, while those of activation and inactivation of kinases becomes
\begin{equation}
   r_{0\rightarrow1}^{A_i} = \frac{c+1}{2} (k + \frac{1}{4}\sum_{j \in n.n.} A_j),
\end{equation}
and
\begin{equation}
    r_{1\rightarrow0}^{A_i} = \frac{\frac{1}{c}+1}{2} (k+1-\frac{1}{4}\sum_{j \in n.n.} A_j).
\end{equation}
The prefactors were chosen to correspond with the case where $p = 0.5$. Because the total rates depend slightly on $p$ when $c\neq1$, we need to choose a specific value of $p$ when running these simulations.

In a traditional Gillespie algorithm, all of these individual propensities are calculated at the beginning of each step of the simulation. Then a single reaction is chosen to take place, with the probability for a given reaction to be chosen is proportional to its propensity. 
The amount of time that passes between each reaction would chosen from an exponential distribution whose timescale is equal to the inverse of the sum of all of the reaction propensities.

In order to speed up simulation times, a form of tau leaping was implemented. Rather than recalculate all the propensities after each reaction, we perform multiple reactions drawing from the same propensity function. The number of reactions performed was chosen to be 
\begin{equation}
    n = 0.1 \textrm{ min}(\sum_j A_j, N - \sum_j A_j),
\end{equation}
rounded up. This guarantees that the total number of active or inactive kinases will not change by more than 10 percent, ensuring that our approximation of a constant propensity function is good for all the reactions. The timestep is then corresponding increased by a factor of $n$.

The units in our simulation are defined so that the spreading rate of activity at $c=1$ is equal to 1. Thus one unit of time in the simulation is equal to 1/$s$ seconds, where $s$ is the spreading rate in units inverse seconds.

Before taking data, if adaptation is occurring the model is run for $3/(k_R+k_B)$ seconds in order to equilibrate. If adaptation is not occurring (i.e $k_R = k_B = 0$) the simulation is run for 100 seconds to equilibrate.

Once the system has equilibrated, we run through whatever experimental protocol we need to. A protocol is defined as a series of ligand concentrations and a length of time the system experiences each ligand concentration in that series. The ligand concentration chosen for equilibration should be the same as the first ligand concentration in the series. The system is run for the specified amount of time at that concentration, then changed to the next one, until the protocol is complete.

\subsection{Resampling}
\begin{figure*}
    \centering
    \includegraphics[width=1\linewidth]{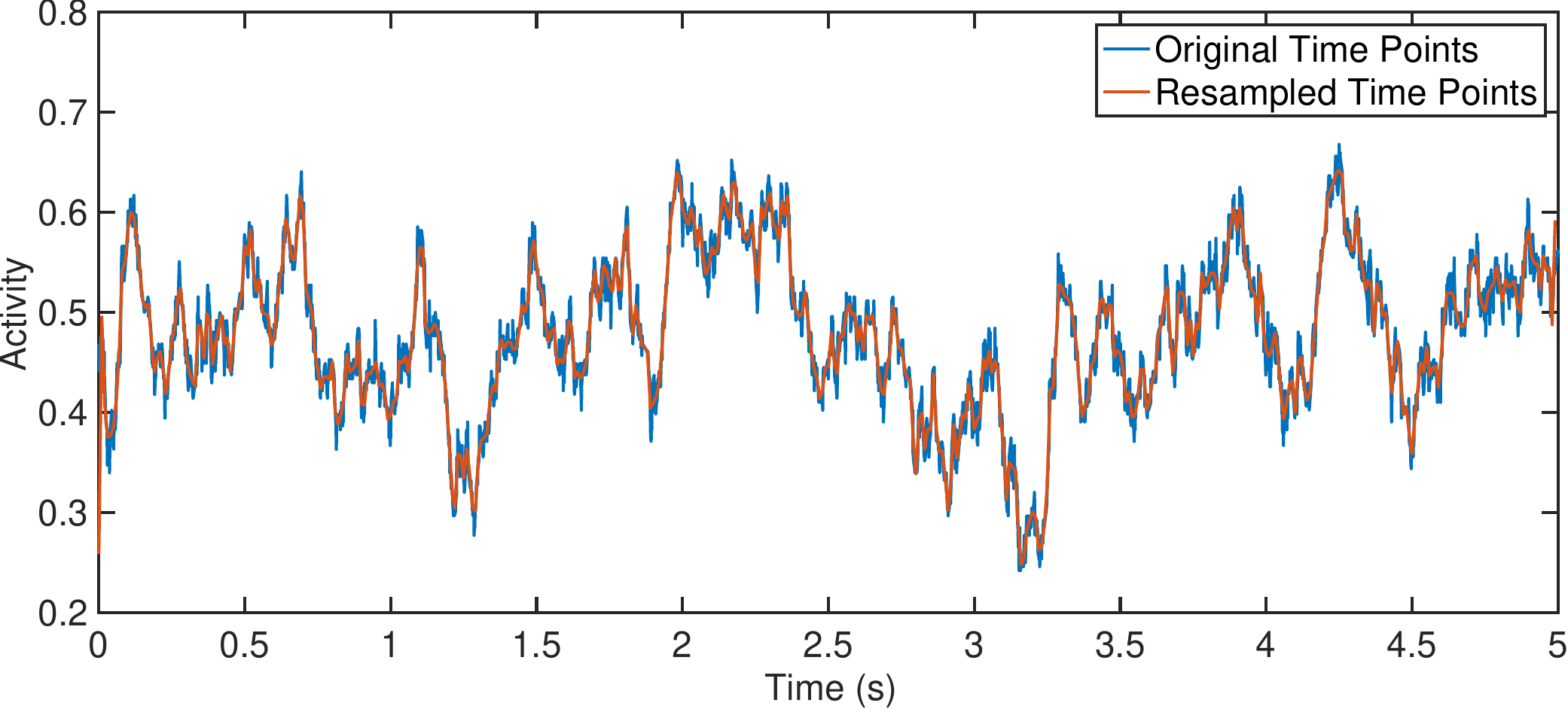}
    \caption{A comparison of original and resampled data. Results are identical, except on timescale less than 0.1 seconds, which are not probe-able using current FRET techniques.}
    \label{fig:sampling}
\end{figure*}

At this point, the model has output a time series of methylation levels and kinase activities, but the spacing between timepoints is non-uniform. For ease of analysis, we then resample the data at 100 Hz using MATLAB's resample function to create a uniformly sampled time series (Fig. \ref{fig:sampling}). From there, it is easy to calculate the power spectrum, or average over many different realizations of the same protocol.

\subsection{Implementing External Noise}
To implement sources of noise in the control parameter that are external to our model, we multiplied $r_{0\rightarrow1}^{A_i}$ by an additional term, $e^{x_\mathrm{OU}}$. ${x_\mathrm{OU}}$ is described by an Ornstein-Uhlenbeck process with a relaxation time $\tau$ and a noise scale $D$. It's value is updated at every time-step of the simulation obeying the following rule:
\begin{equation}
    x_\mathrm{OU}(t+dt) = x_\mathrm{OU}(t)\left(1 - \frac{dt}{\tau}\right) + D \mathrm{N(0,1)} \sqrt{dt}.
\end{equation}
Here, N(0,1) denotes a random variable drawn from a normal distribution with mean 0 and variance 1.

\section{Dependence of Response Times on Receptor State}
\label{sec:response}
We wanted the response speed of the LU model to be roughly independent of ligand concentration once the system reached its adapted steady state activity $\bar{A} = A^*(\bar{c})$. Since the response rate is proportional to sum of the activation and deactivation rates, and ligand concentration only enters into the dynamics through the average receptor state $p$, this is equivalent to wanting the sum of activation and deactivation rates to have minimal dependence on $p$ when $c$ is held constant.

The sum of activation and deactivation rates is
\begin{equation}
    k_{\mathrm{response}} \propto (1+e^{g(m)})p + (1+e^{-g(m)})(1-p).
    \label{eq:Skresponse}
\end{equation}
We also know that
\begin{equation}
    c = \frac{(1+e^{g(m)})p}{(1+e^{-g(m)})(1-p)} = e^{g(m)} \frac{p}{1-p}.
    \label{eq:ScConst}
\end{equation}
Solving for $e^{g(m)}$ in Eq.~\ref{eq:ScConst} and substituting it into Eq.~\ref{eq:Skresponse}, we find
\begin{equation}
    k_{\mathrm{response}} \propto \left(1+\frac{c(1-p)}{p}\right)p + \left(1+ \frac{p}{c(1-p)}\right)(1-p),
\end{equation}
which simplifies to
\begin{equation}
    k_{\mathrm{response}} \propto c(1-p) + (1+\frac{p}{c}).
\end{equation}
If we take the ratio of $k_{\mathrm{response}}$ when $p = 0$ to $k_{\mathrm{response}}$ when $p = 1$, we find that
\begin{equation}
    \frac{k_{\mathrm{response}}(p=0)}{k_{\mathrm{response}}(p=1)} = c.
\end{equation}
That means that, for different values of $p$, and thus ligand concentration, the response rate can change by at most a factor of $c$. Since, as discussed in the main text, $\bar{c}$ is should always be close to 1, the response rate after adaptation should be very close to independent of the ligand concentration adapted to.

\section{Steady State Activity and Approximation}
\label{sec:aproxSS}
The mean field equation for the steady state activity of the LU model is given by Eq. \ref{eq:SA*}. This equation is rather unwieldy, however, so we use Eq. \ref{eq:A*} to approximate it in the main text when we analyze the origins of large gain. Here, we include some additional results supporting the use of the this approximation.

First, we should note that Eq. \ref{eq:A*} is exact in the case where $m = m_{0s}$. It is only when $m\neq m_{0s}$ that it becomes an approximation.

\begin{figure*}
    \centering
    \includegraphics[width=1\linewidth]{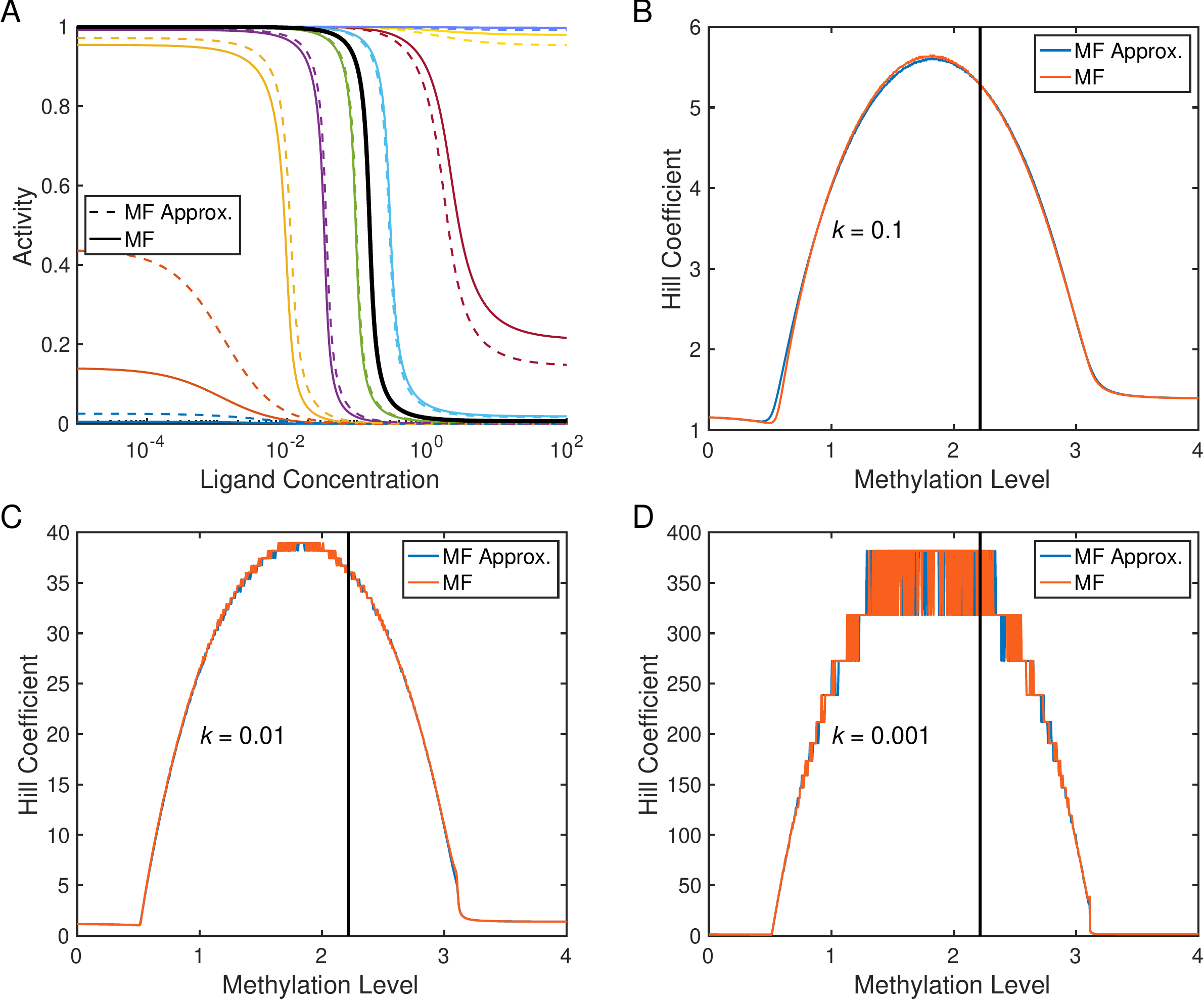}
    \caption{Comparing predictions of exact and approximate equations for the mean-field steady state activity.
    \textbf{A)} Dose response curves for exact and approximate solutions show reasonable good agreement, only diverging significantly for $m$ far from $m_{0s}$. Methylation levels for dose response curves start at $m=0$ in blue at the bottom left, with $m$ increasing by 0.5 for each new color until it reaches purple in the top right at $m=4$. The thick black curve corresponds to $m = m_{0s}$
    \textbf{B-D)} Hill coefficients as a function of $m$ for different values of $k$. The vertical black lines shows the location $m=m_{0s}$. Agreement between the approximate and exact solution is again very good, even when $m$ is far from $m_{0s}$. Discrete step sizes seen at large hill coefficients is an artifact of ligand step-size used in generating dose response curves.}
    \label{fig:MFapprox}
\end{figure*}

To see how the two equations deviated from each other, we compared their predicted dose response curves for various values of $m$ (Fig.~\ref{fig:MFapprox}A). As expected, they are in close agreement near $m=m_{0s}$ and only begin to differ when $m$ changes significantly.

Because we use Eq. \ref{eq:A*} primarily to understand the origins of gain in the LU model, we wanted to specifically check that the gain is well represented by the approximation over a range of $k$ and $m$ values. From both approximate and exact dose response curves, we calculate the Hill coefficients,
\begin{equation}
    h = \frac{\log_{10}(81)}{\log_{10}([L_{90}]/[L_{10}])},   
\end{equation}
where $[L_{90}]$ and $[L_{10}]$ are the ligand concentrations at which a given dose response curve reached 90\% and 10\% of their individual maximum activity. The Hill coefficients show remarkable agreement between the approximation and the exact equations, even for $m$ far from $m_0$ (Fig~\ref{fig:MFapprox}B-D).

\end{document}